\RequirePackage{fix-cm}
\documentclass[smallextended]{svjour3}       % onecolumn (second format)
\smartqed  % flush right qed marks, e.g. at end of proof
\usepackage{graphicx}
\usepackage[colorlinks,citecolor=blue,bookmarks]{hyperref}
\begin{document}

\title{An estimate of Alpha decay half-life from the poles of S-matrix of an exactly solvable potential}%\thanks{Grants or other notes

\titlerunning{An estimate of Alpha decay half-life .......}      % if too long for running head

\author{Swagatika Bhoi, Basudeb Sahu*
        %Second Author %etc.
}


\institute{Swagatika Bhoi \at
              School of Physics, Sambalpur University, Jyoti Vihar, Burla-768019, India.
\\
              \email{bd\_sahu@yahoo.com}           %  \\
           \and
           Basudeb Sahu* \at
              Department of Physics, College of Engineering and Technology
Bhubaneswar-751003, India.
}

\date{Received: date / Accepted: date}

\maketitle

\begin{abstract}
We develop a versatile and analytically solvable potential which fairly reproduces the combined potential of an $\alpha+$nucleus system resulting from both the attractive nuclear and repulsive electrostatic potentials. The potential is expressed in terms of the radial position, mass and proton number of the $\alpha$-particle and the daughter nucleus and certain parameters governing the depth, height and steepness of the barrier. The potential generated is typically a pocket near the origin and a barrier adjacent to it. The Schr\"{o}dinger equation with the above mentioned potential is then solved for the wave function. This potential produces discrete positive energy quasibound state known as resonance state. By matching the wave function and its derivative with the regular Coulomb wave function $F_0$ and irregular Coulomb wave function $G_0$, we obtain the S-matrix analytically. The resonance is obtained from the pole in complex energy plane which gives the width of the corresponding time of decay through the imaginary part of the energy pole position. We make a comparative study of the measured half-lives of various nuclei with the calculated half-lives. The calculated values of half-lives closely match with the corresponding experimental results.
\end{abstract}

\section{Introduction}
Ernest Rutherford's discovery of alpha particle is an incredible milestone in the realm of nuclear physics. It has opened new frontiers of research for full fledged study of alpha decay phenomena. Equally praiseworthy were the discoveries of exotic nuclei including heavy and superheavy nuclei with proton number as large as $118$. Since then $\alpha$-decay has become a topic of intense research and has gained an extraordinary amount of attention for the calculation of half-lives and Q-values.

Earlier findings of decay half-lives have lent support to many theoretical studies namely the relativistic mean field theory, the density dependent M$3$Y (DDM$3$Y) interaction and the Skyrme-Hartree-Fock mean field model \cite{a,b,c,d}. The whole process of decay of $\alpha$-particle from a parent nucleus can be considered as a quantal two body problem comprising of the daughter and the emitted $\alpha$-particle \cite{e}. Indeed, a suitable $\alpha$+daughter nucleus potential can be formulated and its features can be elucidated. The potential chosen can be either phenomenological (eg. Woods-Saxon (WS) shape with adjustable parameters, Ginnochio shape etc.) or it can be formed microscopically by double folding models \cite{f,g}. 

Amongst many of the potentials used so far, Ginocchio potential and Wood-Saxon potentials are the two potentials that have been extensively studied for $\alpha$-decay half-lives. Although the Ginocchio potential is analytically solvable, highly versatile and correctly represent diffuseness still the searching of resonance pole of S-matrix in this case is tedious. On the other hand, searching of poles in Wood-Saxon potential though simple, the potential is not analytically solvable. Therefore, we search for a potential which is analytically solvable and the tracing of poles of the S-matrix is quite easy to estimate the alpha decay half-lives of various exotic nuclei from these poles.

We develop a versatile and analytically solvable potential which fairly reproduces the combined potential of an $\alpha$+nucleus system resulting from both the attractive nuclear and repulsive electrostatic potentials. This potential was used originally by Fiedeldey and Frahn \cite{h}. We have designed a similar potential [Fig. $1$] sans any imaginary part involved. By ascertaining all the parameters, this potential is capable of generating quantal or quasi molecular state and we can easily incorporate $\alpha$-decay process to this quasi molecular decay path.

Many a times WKB method has been applied to get the energies of long-lived states of the effective potential. This semi-classical approximation yields the probability of decay of $\alpha$+daughter nucleus system together with the decay constant. Furthermore, the found out decay constant can be used to get the half-lives. However, this method has serious problems concerning the frequency factor \cite{i}. 

Hence, we sort to an alternative method of finding the half-life by considering the quantum scattering theory where resonance states can be found from the $\alpha$+daughter nucleus two body system. S-matrix theory of potential scattering is used to get the resonance energy. The resonance states are described as poles of the S-matrix falling in the fourth quadrant of the complex-momentum plane. We are therefore left with the complex pole positions which in turn gives both energy and width of the resonance state \cite{g,j,k,l}.

In the present paper, to begin with we develop a potential totally apt for describing the Coulomb nuclear potential of the $\alpha$+nucleus system. Our preferred potential in simple term is a merger of an attractive nuclear part and a repulsive Coulomb part. This merged potential shows a pocket followed by a barrier in its radial variation. We specify the parameters for the depth of the pocket, height and position of the barrier and the steepness of the potential in the interior side. With this potential, Schr\"{o}dinger equation is solved exactly. By matching the wave function and its derivative with the regular Coulomb wave function $F_0$ and irregular Coulomb wave function $G_0$, for partial wave $l=0$, the S-matrix is expressed as a function of the incident energy. In the complex energy or momentum plane, the poles of S-matrix in the fourth quadrant of the momentum plane gives the energy of resonance as well as the width. Subsequently, the width gives the decay time.
 
The above mentioned formulation is applied to a wide range of $\alpha$+nucleus systems including light to heavy to superheavy nuclei for finding out the resonance states and their decay rates. Throughout the paper, we maintain uniformity by changing the value of the parameter $d_1$ which specifically determines the steepness of the inner side of the potential barrier. By doing so, a clear comparison and correlation is done in all the systems considered in the calculation. It is shown that the range of $d_1$ is $2.5 < d_1 < 6.5$.

The paper is organized in the following way. Section II concerns with the formulation of the potential. The experimental results are presented in section III. Summary and conclusions are confronted in section IV.
%\label{intro}

\begin{figure}
%\vspace{-1.7cm}
%\hspace{0.9cm}
\includegraphics[width=1.0\columnwidth,clip=true]{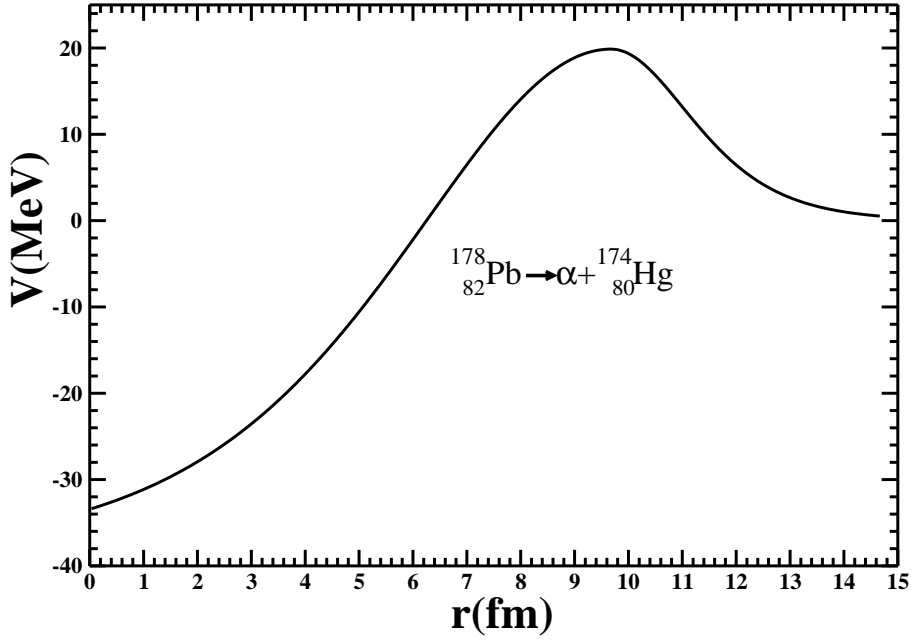}
\label{Fig.1}\vspace{0.2cm} \caption{Plot of the potential V(r) as a function 
of radial distance for the $\alpha+_{80}^{174}$Hg system.}\label{figOne.}
\end{figure}

\section{Formulation}
 We suggest a potential to be a function of radial variable r, which is of the form
\begin{equation}
V(r)=
\left\{
\begin{array}{cl}
V_0[S_1+(S_2-S_1)\rho _1]
&{\hspace{4mm}if\;\;\; r \le R_0,}\\
V_0S_2\rho _2 &{\hspace{4mm}if\;\;\; r \ge R_0,}
\end{array}
\right .
\end{equation}
where $V_0$ is the strength of the potential with value $1$ MeV.
\begin{equation}
\rho_n=\frac{1}{\cosh^2\frac{R_0-r}{d_n}};n=1,2,\nonumber\\
\end{equation}
$d_n$ accounts for the flatness of the barrier, $d_1$ deciding the steepness of the interior side of the barrier whereas the exterior side is judged by $d_2$. $R_0$ is the radial position having value; $R_{0}=r_{0}(A_\alpha ^{1/3}+A_D^{1/3})+2.72$, $r_{0}=0.97$ fm.  Since we are considering the $\alpha$+nucleus system, $A_\alpha$ and $Z_\alpha$ represent the mass and proton number of $\alpha$ particle, $A_D$ and $Z_D$ represent the mass and proton number of the daughter nucleus. Moreover $S_1$ and $S_2$ are the depth and height of the potential, respectively, having values,
\begin{equation}
S_1=-78.75+\frac{3Z_\alpha Z_D e^2}{2R_c},\nonumber \\ 
\end{equation}
\begin{equation}
S_2=\frac{Z_\alpha Z_D e^2}{R_0}(1-\frac{a_g}{R_0}),\nonumber\\
\end{equation}
where $R_c$ is the Coulomb radius parameter; $R_c=r_c(A_\alpha^{1/3}+A_D^{1/3})$, $a_g=1.6$ fm, $r_c=1.2$ fm, $e^2=1.43996$ MeV fm. $r_c$ and $a_g$ are the distance parameters.

With the potential given by (1), the reduced Schr\"{o}dinger equation for the s-wave is written in the dimensionless form as follows:
\begin{eqnarray}
\frac{d^2u_1}{dr^2}+[\kappa^2-k_{0}^{2}S\rho_1]u_1=0,
\end{eqnarray}
where $\kappa^2=k^2-k_{0}^{2} S_1$, $\frac{2m}{\hbar^2} V_0=k_0^2$,$\frac{2m}{\hbar^2} E_{c.m.}=k^2$, $S=S_2-S_1$,
$E_{c.m.}$ refers to the center of mass energy.

The solution $u_{1}(r)$ in the region $r\le R_0$ is given by
\begin{equation}
u_1(r)=A_1 z_1^{{i/2}\kappa d_1} F(a_1,b_1,c_1,z_1) + B_1 z_1^{-i/2\kappa d_1} F(a_1^\prime,b_1^\prime,c_1^\prime,z_1^\prime),\\
\end{equation}
%\end{equation}
\begin{equation}
a_{1}=\frac{1}{2}({\lambda}_{1}+i{\kappa}d_{1}),b_{1}=\frac{1}{2}({1-\lambda}_{1}+i{\kappa}d_{1}),c_{1}=1+i{\kappa}d_{1},\\
\end{equation}
\begin{equation}
a_{1}^{\prime}=\frac{1}{2}({\lambda}_{1}-i{\kappa}d_{1}),b_{1}^{\prime}=\frac{1}{2}({1-\lambda}_{1}-i{\kappa}d_{1}),c_{1}^{\prime}=1-i{\kappa}d_{1},\\
\end{equation}
\begin{equation}
{\lambda}_1=\frac{1}{2}-\frac{1}{2}[1-(2{\kappa}d_1)^2S]^{1/2},\\
\end{equation}
where $z_1=\rho_1(r)$, $F(a,b,c,z)$ is the hyper geometric function and the values of $A_1$ and $B_1$ are decided by using the boundary condition, $u_1(r)=0$ at $r=0$.

Matching the wave function $u_1$ and its derivative with the regular Coulomb wave function $F_0$ and irregular Coulomb wave function $G_0$ at the boundary $r=R_0$, we obtain the S-matrix as
\begin{equation}
S_m=2iC_0+1,
\end{equation}
where
%\begin{equation}
$C_0=\frac{u_1kF_0-u_1^\prime F_0}{u_1^\prime (G_0+iF_0)-u_1k(G_0^\prime 
+iF_0^\prime)}.$%\nonumber
%\end{equation}

The resonance energy is obtained from the poles of S-matrix in the complex momentum plane.
From a resonant pole of the S-matrix expressed by
\begin{equation}
k_P=k_r-ik_i.\nonumber\\
\end{equation}
We obtain the resonance energy (Q value) 
\begin{equation} 
E_P=\frac{\hbar^2}{2m}(k_r^2-k_i^2)\hspace{2mm}MeV,
\end{equation}
and Width
\begin{equation}
\Gamma=\frac{\hbar^2}{2m}  4k_rk_i\hspace{2mm}MeV.
\end{equation}
The time of decay is related to the width as 
\begin{equation}
T_{1/2}=\frac{\hbar}{\Gamma}.
\end{equation}

%%Use table environment for a table in one column
   
%\begin{figure}
%\vspace{-1.7cm}
%\hspace{0.9cm}
%\includegraphics[width=1.0\columnwidth,clip=true]{fig.eps}
%\includegraphics{fig1a.eps}
%\includegraphics{fig1b.eps}
%\label{Fig.1}\vspace{0.2cm} \caption{Plot of the ratio of probability 
%densities $I_1$ and $I_2$ as a function of energy $E_{c.m.}$ for the $\alpha+^{174}_{80}$Hg.The peak position ($E_{c.m.}=7.790$ MeV) represents the resonance energy (Q value).}\label{figOne.}
%\end{figure}
\section{Results and Discussions}
We now apply the formulation as described in section $2$ by taking a potential apt for justifying the $\alpha$+nucleus system. This potential must be compatible with the relativistic mean field theory. For $\alpha+^{174}_{80}$Hg system we reproduce the same potential with our analytically solvable potential expressed by (1) by assigning the values of the parameters $R_0=9.6749$ fm, $S_1=-38.5841$ MeV, $S_2=19.8752$ MeV, $d_1=5.13240$, $d_2=2$. This potential for the system is shown in Fig. 1. Using this potential we obtain the resonance energy, Q-value and the half-life from the resonant pole of S-matrix given by (10). We now sagaciously change the values of the potential parameter $d_1$ and apply the same potential formulation to a wide range of systems spanning the light, heavy and superheavy nuclei to get the half lives ready for close comparison with the corresponding experimental values. For good agreement of the experimental values we keep all the parameters except changing one parameter concerning the steepness of the potential barrier in the interior side. Keen observation on the outcome reveal that the value of $d_1$ deviate in the range $2.5$ to $6.5$.

For a given system we vary the value of $d_1$ and locate a pole to keep the energy equal to the experimental Q-value. The width of this pole gives us the $T_{1/2}$ through (14) and we represent this by $T_{1/2}^{calt.}$. We now jot the trio i.e $Q^{expt}$, $T^{expt.}$ and $T^{calt.}$ in Table 1 and Table 2 for a wide range of $\alpha$+nucleus systems. The results in Table 1 and Table 2 show that for the whole range of $\alpha$-decaying nuclei system the experimental and calculated half-lives closely match. We also confirm the close agreement for the case of different isotopes of Po nucleus which is shown in Fig. 3. 

We hand-pick a group of systems by going down the tabulation (i.e $^{146}_{62}$
Sm, $^{150}_{64}$Gd, $^{154}_{66}$Dy, $^{190}_{78}$Pt, $^{210}_{82}$Pb, $^{226}_{88}$Ra, $^{232}_{90}$Th, $^{238}_{92}$U, $^{244}_{94}$Pu, $^{250}_{96}$Cm). A glimpse on the results of the systems affirms that they have very large experimental decay time, $T_{1/2}^{expt.}$.  Let us consider the case of $^{146}_{62}$Sm for which $T_{1/2}^{expt.}=2.2\times10^{15}$ s. Our calculated value of $T_{1/2}^{calt.}=3.73\times10^{15}$ s which is very close to the above experimental value. Among all the systems chosen in this paper, the longest half-life is seen in case of $^{238}_{92}$U where $T_{1/2}^{expt.}=1.4\times10^{17}$ s and the corresponding width $\Gamma ^{expt.}=3.25\times10^{-39}$ MeV for Q value of $4.270$ MeV. Our calculated results show $T_{1/2}^{calt.}=5.59\times10^{17}$ s with $\Gamma ^{calt.}=8.15\times10^{-40}$ MeV. Comprehensive results are obtained for a series of nuclei and the experimental and calculated results match closely for very small decay time as well as for very long decay time.

\begin{table}
\caption{\label{tab:table1}Comparison between the experimental $\alpha$-decay half-lives \cite{d} and results of the present calculation. The value of the parameter $d_2$ is kept constant throughout i.e. $d_2=2$.}
\renewcommand{\tabcolsep}{0.5cm}
\renewcommand{\arraystretch}{1.6}
\footnotesize
\begin{tabular}{c c c c c}
\hline
\hline
Nucleus&$d_1$&Q$_\alpha^{expt.}$(MeV)& $T_{1/2}^{expt.}(s)$& $T_{1/2}^{calt.}(s)$
%%Nuclei&\multicolumn{3}{c|}{BE}&\multicolumn{3}{c|}{$\beta_2$}&\multicolumn{3}{c|}{$r_c$}&\multicolumn{3}{c|}{$r_t$}&\multicolumn{2}{c}{FRDM}\\
%%\cline{2-4} \cline{5-7} \cline{8-10} \cline{11-13} \cline{14-15}
%%&sph.&prol.&obl.&sph.&prol.&obl.&sph.&prol&obl.&sph.&prol.&obl.&BE&$\beta_2$\\
\\ \hline\hline
$^{106}_{52}$Te&4.90420& 4.290 & 8.0 $	\times$10$^{-5}$ & 4.02 $\times$ 10$^{-5}$ 
 \\
$^{108}_{52}$Te&4.81479& 3.420 & 4.3 $	\times$10$^{0}$ & 2.13 $\times$10$^{0}$
  \\
$^{110}_{52}$Te&4.75180& 2.699 & 6.2 $	\times$10$^{5}$& 8.39 $	\times$10$^{5}$
 \\
$^{112}_{54}$Xe&4.11499& 3.330 & 3.9 $	\times$10$^{0}$& 1.52 $	\times$10$^{0}$
 \\
$^{146}_{62}$Sm&4.90780& 2.528 & 2.2 $	\times$10$^{15}$& 3.73 $\times$10$^{15}$
  \\
$^{148}_{64}$Gd&4.94120& 3.271 & 2.2 $	\times$10$^{9}$& 2.35 $	\times$10$^{9}$
  \\
$^{150}_{64}$Gd&4.907800& 2.808 & 5.6 $\times$10$^{13}$& 7.87 $\times$10$^{13}$
  \\
$^{150}_{66}$Dy&6.47199& 4.351 & 1.2 $	\times$10$^{3}$& 4.87 $\times$10$^{3}$
 \\
$^{152}_{66}$Dy&3.72760& 3.726 & 8.6 $	\times$10$^{6}$& 2.58 $\times$10$^{6}$
 \\
$^{154}_{66}$Dy&4.88380& 2.945 & 9.5 $	\times$10$^{13}$& 8.16 $\times$10$^{13}$
 \\
$^{152}_{68}$Er&6.48559& 4.934 & 1.1 $	\times$10$^{1}$& 4.75 $\times$10$^{1}$
 \\
$^{156}_{68}$Er&3.64480& 3.483 & 6.7 $	\times$10$^{9}$& 3.87 $\times$10$^{9}$
 \\
$^{154}_{70}$Yb&5.04180& 5.474 & 4.4 $	\times$10$^{-1}$& 2.89 $\times$10$^{-1}$
 \\
$^{158}_{70}$Yb&4.90700& 4.170 & 4.3 $	\times$10$^{6}$& 2.67 $\times$10$^{6}$
 \\
$^{156}_{72}$Hf&5.04580& 6.028 & 2.4 $	\times$10$^{-2}$& 1.45 $\times$10$^{-2}$
 \\
$^{162}_{72}$Hf&3.66560& 4.416 & 4.9 $	\times$10$^{5}$ & 2.00 $\times$10$^{5}$
 \\
$^{158}_{74}$W&6.51280& 6.613 & 1.3 $	\times$10$^{-3}$ & 4.36 $\times$10$^{-3}$
   \\
$^{168}_{74}$W&4.90140& 4.500 & 1.6 $	\times$10$^{6}$& 3.83 $\times$10$^{6}$
 \\
$^{164}_{76}$Os&5.07320& 6.767 & 2.1 $	\times$10$^{-3}$& 1.67 $\times$10$^{-3}$
 \\
$^{166}_{78}$Pt&5.01939& 6.990 & 2.0 $	\times$10$^{-3}$& 2.14 $\times$10$^{-3}$
  \\
$^{190}_{78}$Pt&4.84759& 3.252 & 2.0 $	\times$10$^{19}$& 1.44 $\times$10$^{19}$
  \\
$^{172}_{80}$Hg&5.10033& 7.524 & 2.3 $	\times$10$^{-4}$& 2.51 $\times$10$^{-4}$
  \\
$^{188}_{80}$Hg&4.92299& 4.703 & 5.2 $	\times$10$^{8}$& 4.09 $\times$10$^{8}$
 \\
$^{178}_{82}$Pb&5.13240& 7.790 & 2.3 $	\times$10$^{-4}$& 2.23 $\times$10$^{-4}$
 \\
$^{180}_{82}$Pb&5.10599& 7.419 & 4.2 $	\times$10$^{-3}$& 3.10 $\times$10$^{-3}$
 \\
$^{182}_{82}$Pb&5.08279& 7.066 & 5.5 $	\times$10$^{-2}$& 4.22 $\times$10$^{-2}$
 \\
$^{184}_{82}$Pb&5.06999& 6.774 & 6.1 $	\times$10$^{-1}$& 4.67 $\times$10$^{-1}$
 \\
$^{186}_{82}$Pb&5.05500& 6.470 & 1.2 $	\times$10$^{1}$& 6.81 $\times$10$^{0}$
 \\
$^{188}_{82}$Pb&5.03020& 6.109 & 2.8 $	\times$10$^{2}$& 1.94 $\times$10$^{2}$
 \\
$^{190}_{82}$Pb&4.99679& 5.697 & 1.8 $	\times$10$^{4}$& 1.49 $\times$10$^{4}$
 \\
$^{192}_{82}$Pb&4.95260& 5.221 & 3.5 $	\times$10$^{6}$ & 4.42 $\times$10$^{6}$
 \\
$^{194}_{82}$Pb&4.90780& 4.738 & 8.8 $	\times$10$^{9}$ & 3.06 $\times$10$^{9}$
   \\
$^{210}_{82}$Pb&5.01999& 3.792 & 3.7 $	\times$10$^{16}$& 3.7 $	\times$10$^{16}$
 \\
$^{190}_{84}$Po&5.21159& 7.693 & 2.5 $	\times$10$^{-3}$& 2.21 $\times$10$^{-3}$
 \\
$^{192}_{84}$Po&5.18260& 7.320 & 3.4 $	\times$10$^{-2}$& 3.54 $\times$10$^{-2}$
  \\
$^{194}_{84}$Po&5.16059& 6.987 & 3.9 $	\times$10$^{-1}$& 4.68 $\times$10$^{-1}$
  \\
$^{196}_{84}$Po&5.13940& 6.658 & 5.7 $	\times$10$^{0}$& 8.01 $	\times$10$^{0}$
\\

\hline\hline
\end{tabular}
\end{table}

\begin{table}
\caption{\label{tab:table1}Comparison between the experimental $\alpha$-decay half-lives \cite{d} and results of the present calculation. The value of the parameter $d_2$ is kept constant throughout i.e. $d_2=2$.}
\renewcommand{\tabcolsep}{0.5cm}
\renewcommand{\arraystretch}{1.6}
\footnotesize
\begin{tabular}{c c c c c}
\hline
\hline
Nucleus&$d_1$&Q$_\alpha^{expt.}$(MeV)& $T_{1/2}^{expt.}(s)$& $T_{1/2}^{calt.}(s)$
%%Nuclei&\multicolumn{3}{c|}{BE}&\multicolumn{3}{c|}{$\beta_2$}&\multicolumn{3}{c|}{$r_c$}&\multicolumn{3}{c|}{$r_t$}&\multicolumn{2}{c}{FRDM}\\
%%\cline{2-4} \cline{5-7} \cline{8-10} \cline{11-13} \cline{14-15}
%%&sph.&prol.&obl.&sph.&prol.&obl.&sph.&prol&obl.&sph.&prol.&obl.&BE&$\beta_2$\\
\\ \hline\hline
$^{198}_{84}$Po&5.11460& 6.309 & 1.9 $\times$10$^{2}$& 2.10 $\times$10$^{2}$
 \\
$^{200}_{84}$Po&5.09320& 5.981 & 6.2 $	\times$10$^{3}$& 5.44 $	\times$10$^{3}$
  \\
$^{202}_{84}$Po&5.08020& 5.701 & 1.4 $	\times$10$^{5}$& 1.17 $	\times$10$^{5}$
 \\
$^{204}_{84}$Po&5.07760& 5.485 & 1.9 $	\times$10$^{6}$& 1.51 $	\times$10$^{6}$
 \\
$^{206}_{84}$Po&6.47139& 5.327 & 1.4 $\times$10$^{7}$& 6.98 $	\times$10$^{7}$
  \\
$^{208}_{84}$Po&5.09919& 5.215 & 9.1 $	\times$10$^{7}$& 4.18 $	\times$10$^{7}$
  \\
$^{210}_{84}$Po&6.56080& 5.407 & 1.2 $\times$10$^{7}$& 2.29 $	\times$10$^{7}$
  \\
$^{212}_{84}$Po&5.85500& 8.954 & 3.0 $\times$10$^{-7}$ & 4.64 $\times$10$^{-7}$
 \\
$^{214}_{84}$Po&4.37020& 7.833 & 1.6 $\times$10$^{-4}$ & 1.17 $\times$10$^{-4}$
 \\
$^{216}_{84}$Po&4.25500& 6.906 & 1.5 $\times$10$^{-1}$& 1.2 $	\times$10$^{-1}$
 \\
$^{218}_{84}$Po&4.16359& 6.115 & 1.9 $\times$10$^{2}$& 1.89 $\times$10$^{2}$
 \\
$^{198}_{86}$Rn&3.90799& 7.349 & 6.6 $\times$10$^{-2}$& 3.43 $\times$10$^{-2}$
 \\
$^{222}_{86}$Rn&4.02780& 5.590 & 3.3 $\times$10$^{5}$& 5.39 $\times$10$^{5}$
 \\
$^{206}_{88}$Ra&3.93280& 7.415 & 2.4 $\times$10$^{-1}$& 1.11 $\times$10$^{-1}$
 \\
$^{226}_{88}$Ra&2.72480& 4.871 & 5.0 $\times$10$^{10}$& 1.57 $\times$10$^{10}$
 \\
 $^{214}_{90}$Th&4.00759& 7.827 & 8.7 $\times$10$^{-2}$& 2.74 $\times$10$^{-2}$
 \\
$^{232}_{90}$Th&2.59220& 4.082 & 4.4 $\times$10$^{17}$& 2.23 $\times$10$^{17}$
   \\
$^{222}_{92}$U&4.27100& 9.430 & 1.5 $\times$10$^{-6}$& 3.80 $\times$10$^{-6}$
 \\
$^{238}_{92}$U&3.71900& 4.270 & 1.4 $\times$10$^{17}$& 5.59 $\times$10$^{17}$
 \\
$^{228}_{94}$Pu&5.26060& 7.940 & 2.1 $\times$10$^{0}$& 1.94 $\times$10$^{0}$
  \\
$^{244}_{94}$Pu&3.74699& 4.666 & 2.5 $\times$10$^{15}$& 4.30 $\times$10$^{-3}$
  \\
$^{238}_{96}$Cm&2.69679& 6.670 & 7.9 $\times$10$^{4}$& 2.44 $\times$10$^{4}$
  \\
$^{250}_{96}$Cm&2.65660& 5.169 & 1.5 $\times$10$^{12}$ & 2.39 $\times$10$^{12}$ 
 \\
$^{240}_{98}$Cf&3.91659& 7.711 & 4.1 $\times$10$^{1}$ & 6.45 $\times$10$^{1}$
 \\
$^{254}_{98}$Cf&3.84459& 5.927 & 1.7 $\times$10$^{9}$& 4.56 $\times$10$^{9}$
 \\
$^{246}_{100}$Fm&3.99060& 8.377 & 1.3 $\times$10$^{0}$& 1.92 $\times$10$^{0}$
 \\
$^{256}_{100}$Fm&3.92460& 7.027 & 1.2 $\times$10$^{5}$& 1.82 $\times$10$^{5}$
 \\
$^{252}_{102}$No&3.98620& 8.548 & 4.1 $\times$10$^{0}$& 2.89 $\times$10$^{0}$
 \\
$^{256}_{102}$No&4.04999& 8.581 & 2.9 $\times$10$^{0}$& 2.09 $\times$10$^{0}$
 \\
$^{254}_{104}$Rf&4.00059& 9.210 & 1.7 $\times$10$^{-1}$& 1.33 $\times$10$^{-1}$
 \\
$^{258}_{104}$Rf&4.05620& 9.190 & 1.1 $\times$10$^{-1}$& 1.41 $\times$10$^{-1}$ 
 \\
$^{260}_{106}$Sg&5.33759& 9.901 & 1.2 $\times$10$^{-2}$& 4.63 $\times$10$^{-2}$
   \\
$^{264}_{108}$Hs&5.39579& 10.591 & 1.1 $\times$10$^{-3}$& 3.39 $\times$10$^{-3}$
 \\
$^{270}_{110}$Ds&4.18099& 11.117 & 2.1 $\times$10$^{-4}$& 1.16 $\times$10$^{-4}$
 \\
$^{286}_{114}$Fl&4.03920& 10.370 & 1.4 $\times$10$^{-1}$& 1.33 $\times$10$^{-1}$
  \\
$^{290}_{116}$Lv&2.89960& 10.990 & 8.0 $\times$10$^{-3}$& 2.24 $\times$10$^{-3}$
  \\
$^{294}_{118}$Lv&5.40980& 11.810 & 1.4 $\times$10$^{-3}$& 4.30
 $\times$10$^{-3}$
\\

\hline\hline
\end{tabular}
\end{table}
\begin{figure}
%\vspace{-1.7cm}
%\hspace{0.9cm}
\includegraphics[width=1.0\columnwidth,clip=true]{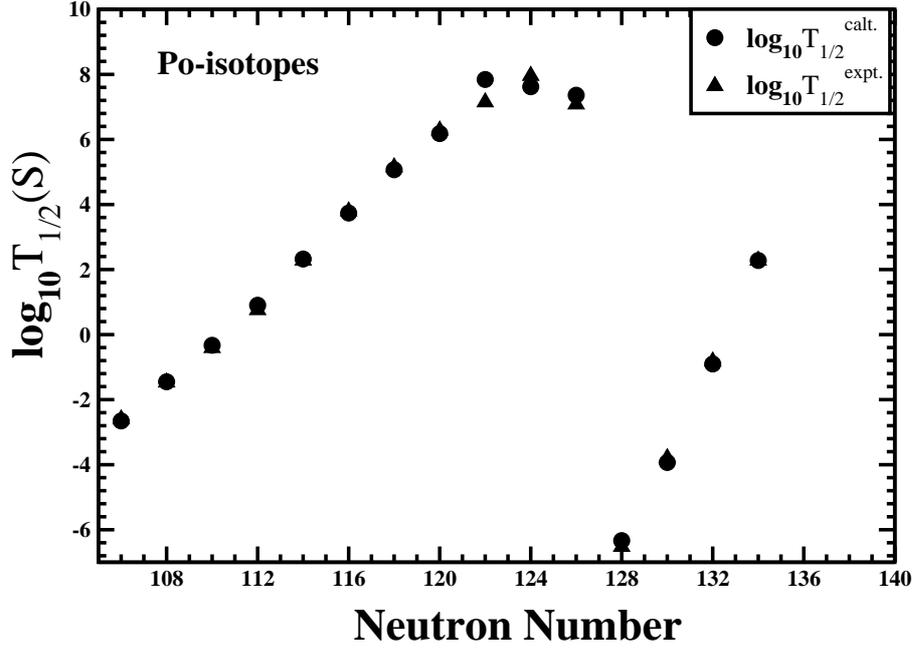}
\label{Fig.1}\vspace{0.2cm} \caption{$\alpha$-decay half-lives for different isotopes of Po nucleus. Solid dots represent the calculated results and the solid triangles represent the experimental results.}\label{figOne.}
\end{figure}

\section{Summary and Conclusion}
An analytical expression for a potential is developed to simulate the combined effect of the attractive nuclear and repulsive electrostatic force. We have given expressions for the potential which is analytically solvable. Using this potential the s-wave Schr\"{o}dinger equation is solved and the S-matrix is obtained by matching the wave function and its derivative with the Coulomb wave function outside. From the expressions of S-matrix the pole representing the resonance state is searched. From this pole we obtain the Q-value and the width or decay half life of the resonance state.

Results of $\alpha$-decay half-lives have been calculated for various nuclei starting from a nucleus with $A=106(Z=52)$ to a nucleus with $A=294(Z=118)$. A thorough comparison of these calculated results with the corresponding experimental values shows that the calculated half-lives are compatible with the experimental half-lives. Thus, this formulation can be used in explaining the half-lives of $\alpha$-decay of any nuclei that include light, heavy and super heavy showing decay time from very small ($10^{-4}$ s) to very large decay time of the order of $10^{17}$ s.

\section{Acknowledgement}
We gratefully acknowledge the computing and library facilities extended by the Institute of Physics, Bhubaneswar.

\end{document}